\documentclass[aps,prd,onecolumn,eqsecnum,amsmath,nofootinbib,preprintnumbers,superscriptaddress]{revtex4-2}

\usepackage[dvipsnames]{xcolor}
\usepackage{color,graphicx,float}
\usepackage{amsfonts,amssymb,mathrsfs,times}
\usepackage{bm}
\usepackage{multirow}
\usepackage{mathtools}
\usepackage{dsfont}
\usepackage{setspace}
\usepackage[colorlinks,linkcolor=blue,anchorcolor=blue,citecolor=blue]{hyperref}
\usepackage{ulem}
\usepackage{subcaption}
\usepackage{amsthm}

\usepackage[justification=raggedright,singlelinecheck=false]{caption}

\begin{document}
\title{Exceptional point excitation by a plunging particle}
\date{\today}

\author{Xia-Yuan Liu}
\email{liuxiayuan@mail.ustc.edu.cn}
\affiliation{Interdisciplinary Center for Theoretical Study and Department of Modern Physics,\\
University of Science and Technology of China, Hefei, Anhui 230026, China}

\author{Liang-Bi Wu}
\email[corresponding author: ]{wulb@ucas.ac.cn}
\affiliation{School of Fundamental Physics and Mathematical Sciences, Hangzhou Institute for Advanced Study, UCAS, Hangzhou, Zhejiang 310024, China}


\begin{abstract}
We investigate the excitation of a second-order exceptional point (EP) in the quasinormal mode (QNM) spectrum by a plunging particle. Starting from a gravitational action, we derive the master equation with a Gaussian bump modified potential. For the odd-parity mode $(\ell,m)=(2,1)$, we compute the waveform numerically and derive an analytic expression for the EP contribution to the ringdown waveform directly from the particle source. Comparison with the numerical waveform shows that the particle excites the EP response, which, together with its mirror contribution, dominates the intermediate ringdown stage. We use waveform fitting and Bayesian analysis to assess the limitations of conventional QNM templates for EP ringdown. In the fitting, the EP template recovers the complex frequency $\omega$ more accurately than the ordinary template. In Bayesian analysis, at a signal-to-noise ratio (SNR) of $\rho=100$, the ordinary template yields a narrow frequency posterior whose $90\%$ credible region excludes the true EP frequency, whereas the broader EP-template region contains it. The Bayes factor nevertheless favors the ordinary template at lower SNRs and the EP template only at higher SNRs. These results show that in the EP case a narrow posterior and favorable Bayesian evidence do not guarantee accurate frequency inference when a conventional template omits the characteristic EP ringdown structure.
\end{abstract}

\maketitle

\section{Introduction}

Since LIGO's first direct detection of gravitational waves (GW) from a binary black hole merger in 2015, gravitational-wave astronomy has advanced rapidly \cite{LIGOScientific:2016aoc}. Ground-based detectors have now observed mergers involving stellar-mass black holes and neutron stars \cite{LIGOScientific:2017vwq}. Future space-based detectors, including LISA, Taiji, and TianQin, will probe the millihertz frequency band, with massive black hole binaries, Galactic compact binaries, and extreme-mass-ratio inspirals (EMRIs) among their main targets \cite{LISA:2024hlh,Ruan:2018tsw,TianQin:2020hid}. At nanohertz frequencies, pulsar timing arrays have also found evidence for a stochastic gravitational-wave background, possibly produced predominantly by a population of supermassive black hole binaries \cite{NANOGrav:2023gor,EPTA:2023fyk,NANOGrav:2023hfp}. Accurate theoretical waveforms are essential for interpreting these signals~\cite{Cutler:2007mi}. Black hole perturbation theory provides a framework for describing both the ringdown of a black hole formed in a merger and the gravitational perturbations generated by a small compact object in an extreme-mass-ratio system~\cite{Regge:1957td,Pound:2021qin}. The oscillation and damping of black hole ringdown are characterized by quasinormal modes (QNMs)~\cite{Kokkotas:1999bd,Konoplya:2011qq}, whose complex frequencies form the basis of black hole spectroscopy~\cite{Berti:2009kk,Dreyer:2003bv,Berti:2005ys,Berti:2025hly}.

In general relativity, the QNM spectra of isolated Schwarzschild and Kerr black holes are determined by their mass and spin~\cite{Berti:2009kk}. However, surrounding matter or modifications of gravity theory alter the QNM spectrum significantly~\cite{Barausse:2014tra,Cardoso:2018ptl,Jaramillo:2020tuu} due to the non-Hermitian nature. Recent studies have shown that even small localized deformations of the effective potential can produce large shifts in certain QNM spectra~\cite{Cheung:2021bol,Yang:2024vor}. As the system parameters vary, QNM branches may exhibit avoided crossings~\cite{Dias:2022oqm,Motohashi:2024fwt,Lo:2025njp,Takahashi:2025uwo,Jenks:2026tae,Takahashi:2026daf,Kubota:2025hjk}. At particular parameter values, two QNMs can also coalesce in both their frequencies and eigenfunctions, forming an exceptional point (EP)~\cite{Heiss:2012dx,Ashida:2020dkc}. For the black hole perturbation theory, EPs are investigated in various models~\cite{Cavalcante:2024swt,Cavalcante:2024kmy,Cavalcante:2025abr,Oshita:2025ibu,Rossi:2026als,Nakamoto:2026lyo,Cavalcante:2026vgr,Arcos:2026eis,OuldElHadj:2026vym,Yang:2025dbn,Cao:2025afs,Hu:2025efp,OuldElHadj:2026esa,Morikawa:2026fzn,PanossoMacedo:2025xnf,Cheng:2026gxu,Wu:2025wbp,Long:2026vpw,Wu:2026tfx,Hu:2026tgi,Destounis:2026qne,Destounis:2026xvd,Zhang:2026lyu,Macedo:2025ep,Blazquez-Salcedo:2026wbi}. In a larger parameter space, these degeneracies can form continuous exceptional lines (ELs)~\cite{Cao:2025afs,Cavalcante:2026vgr,Wu:2026tfx,Nakamoto:2026lyo}. Such ELs exhibit an intrinsically directional spectral response~\cite{Cao:2025afs,Wu:2026tfx}. An EP also changes the form of the associated ringdown contribution. In Ref. \cite{Yang:2025dbn}, the authors studied a Schwarzschild model with a Gaussian deformation of the axial RW potential. By tuning the amplitude and position of the deformation, they found a second-order EP. The corresponding double pole of the frequency-domain Green function gives rise to a ringdown contribution of the form $(A+Bu)\mathrm{e}^{-\mathrm{i}\omega_\star u}$, where $u$ is retarded time~\cite{Yang:2025dbn,Macedo:2025ep}. Using numerical evolutions of Gaussian initial perturbations, they showed that conventional QNM templates can yield significantly biased estimates of the complex frequency, whereas including the linear-in-time factor substantially improves frequency recovery.

Previous studies have examined the EP ringdown response using freely prescribed initial perturbations~\cite{Yang:2025dbn,Macedo:2025ep,Cheng:2026gxu,Zhang:2026lyu,OuldElHadj:2026vym}, but it remains unclear how the characteristic EP features manifest in gravitational waves generated by a physical source. To address this question, we consider a point particle plunging into the black hole, with a source term determined by its motion~\cite{Hadar:2009ip,Silva:2023cer,DeAmicis:2025xuh}, and investigate whether the characteristic EP response is excited and becomes apparent in the resulting gravitational waveform. A complementary motivation comes from EMRI, an important area of gravitational-wave research~\cite{Babak:2017tow,Barack:2018yvs}. From this perspective, our model describes an idealized EMRI plunge in which the central black hole's QNM spectrum contains an EP, allowing us to examine its effects on the plunge and subsequent ringdown waveform. Related challenges in resolving nearly degenerate modes have been examined with Bayesian inference~\cite{Imafuku:2026rpn}.

Starting from an action, we derive a perturbation equation with a Gaussian bump in the effective potential, providing a theoretical interpretation of this previously phenomenological class of models~\cite{Cheung:2021bol,DellaRocca:2026zym}. We then numerically solve the homogeneous perturbation equation, examine the properties of its solutions, and analytically derive the EP ringdown waveform from the particle source. By studying the QNM spectrum near the EP and the full waveform generated by a plunging particle, we show that the EP response is excited by the particle and, together with its mirror contribution, dominates the intermediate ringdown stage. We subsequently investigate its implications for frequency inference through waveform fitting and Bayesian analysis. We find that a conventional QNM template will yield a biased complex frequency. At the finite signal-to-noise ratios (SNRs) considered, its frequency posterior interval is narrow but excludes the true value, whereas the broader interval obtained with the EP template contains it. A comparison of Bayesian evidences further shows that the data favor the conventional template at lower SNRs and the EP template at higher SNRs. This means that, without prior knowledge that the system is at an EP, we may select the conventional template based on the Bayesian evidence at lower SNRs and interpret its good waveform fit and narrow posterior interval as a reliable frequency measurement, even though that interval excludes the true value. Our study therefore highlights a potential pitfall in gravitational-wave inference: a good waveform fit and Bayesian evidence favoring the adopted template do not necessarily guarantee reliable results.

The remainder of this work is organized as follows. Section \ref{sec:theory} specifies the modified potential, plunge trajectory, and particle source. Section \ref{sec:response} presents the homogeneous solutions and the frequency-domain Green function, and determines the QNM spectrum and the EP. Section \ref{sec:results} derives the analytic EP response from the particle source, calculates the numerical waveforms, and examines QNM frequency inference through waveform fitting and Bayesian analysis. Section \ref{sec:conclusions} summarizes our findings and discusses future directions. Appendix \ref{app:reduced_action} derives the perturbation equation with the Gaussian bump modified potential and its particle source term from an action, and establishes the relation between the RW function and the strain.

\section{Physical model and plunging source}
\label{sec:theory}
We consider a point particle of mass $\mu\ll M$ plunging into a Schwarzschild black hole of mass $M$. The background geometry and the leading-order trajectory are those of Schwarzschild, while the axial quadrupolar perturbation propagates in a modified Regge--Wheeler (RW) potential.  We use units $G=c=1$, and the background metric is
\begin{eqnarray}\label{eq:schwarzschild}
    \mathrm{d}s^2=-f(r)\mathrm{d}t^2+f^{-1}(r)\mathrm{d}r^2+r^2\mathrm{d}\Omega^2\, ,\qquad f(r)=1-\frac{2M}{r}\, ,
\end{eqnarray}
and the tortoise coordinate $x$ is
\begin{eqnarray}\label{eq:tortoise}
    x=r+2M\ln\Big(\frac{r-2M}{M}\Big)\, ,\qquad\frac{\mathrm{d}x}{\mathrm{d}r}=f(r)^{-1}\, .
\end{eqnarray}
For odd-parity perturbations we use the RW master function $\Psi_{\ell m}$ \cite{Regge:1957td} in the normalization of Refs. \cite{Martel:2003jj,Martel:2005ir}, which satisfies
\begin{eqnarray}\label{eq:master}
    \Big[-\partial_t^2+\partial_x^2-V_\ell(x)\Big]\Psi_{\ell m}(t,x)=S_{\ell m}(t,x)\, .
\end{eqnarray}
Following Ref. \cite{Yang:2025dbn}, we add a Gaussian bump
to the $\ell=2$ potential only,
\begin{eqnarray}\label{eq:potential}
    V_2(x)=V_2^{\rm RW}(x)+\frac{\epsilon}{M^2}\exp\Big[-\Big(\frac{x-d}{M}\Big)^2\Big]\, ,\quad\text{with}\quad V_\ell^{\rm RW}=f(r)\Big[\frac{\ell(\ell+1)}{r^2}-\frac{6M}{r^3}\Big]\, .
\end{eqnarray}
Here $\epsilon$ is the dimensionless amplitude, $d$ is the center of the bump in the tortoise coordinate, and the width parameter is fixed to $M$.

In black hole perturbation theory, adding a small localized bump to the RW effective potential is a commonly used phenomenological approach to modeling the effects of surrounding matter or other localized structures on gravitational perturbations. Cheung \textit{et al.} showed that even a weak localized perturbation, modeled by a P\"{o}schl--Teller or Gaussian bump, can produce substantial shifts in the QNM spectra~\cite{Cheung:2021bol}. Subsequently, Yang \textit{et al.} demonstrated that tuning the amplitude and position of a Gaussian bump can cause two QNMs to coalesce at an exceptional point (EP)~\cite{Yang:2025dbn}. However, such potential deformations are usually prescribed phenomenologically rather than derived from a specific matter model or a gravitational action. Recent work has also shown that, in several commonly studied fluid models, describing environmental effects solely through a localized potential bump may be incompatible with the regularity and causality requirements of the matter distribution~\cite{DellaRocca:2026zym}.

To provide a covariant interpretation of the Gaussian bump, we consider a gravitational theory containing a curvature coupling $\mathcal A(\mathcal K)\mathcal P^2$, where $\mathcal K$ is the Kretschmann scalar and $\mathcal P$ is the Pontryagin pseudoscalar~\cite{Cardoso:2018ptl,Nakashi:2025fbr}. Since $\mathcal P$ vanishes on the Schwarzschild background, this coupling leaves the background geometry unchanged while modifying the axial gravitational perturbation equation. In this model, the background spacetime and the particle trajectory at leading order remain those of Schwarzschild, while the additional curvature coupling modifies the effective potential for axial perturbations. The same coupling also introduces a small correction to the point particle source in the RW equation. The corresponding detailed derivation is given in Appendix \ref{app:reduced_action}.

To specify the spectrum structure, we first consider the homogeneous problem, $S_{\ell m}=0$. With the Fourier convention $\widetilde\Psi(\omega)=\int \mathrm{d}t\,\mathrm{e}^{\mathrm{i}\omega t}\Psi(t)$, stable QNMs have $\operatorname{Im}\omega<0$, and the radial equation $(\partial_x^2+\omega^2-V_\ell)X_\ell(\omega,x)=0$ has solutions
\begin{eqnarray}\label{eq:jost}
    X_\ell^{\rm in}(\omega,x)\sim \mathrm{e}^{-\mathrm{i}\omega x}\quad (x\to-\infty)\, ,\qquad X_\ell^{\rm up}(\omega,x)\sim \mathrm{e}^{\mathrm{i}\omega x} \quad (x\to+\infty)\, ,
\end{eqnarray}
so that $X_\ell^{\rm in}$ is purely ingoing at the horizon and $X_\ell^{\rm up}$ purely outgoing at infinity. Their Wronskian,
\begin{eqnarray}\label{eq:wronskian}
    W_\ell(\omega)=X_\ell^{\rm in}\partial_xX_\ell^{\rm up}-(\partial_xX_\ell^{\rm in})X_\ell^{\rm up}\, ,
\end{eqnarray}
is independent of $x$, and its zeros are the QNM spectra. For suitable $(\epsilon,d)$, two QNMs and their radial eigenfunctions coalesce at a second-order EP,
\begin{eqnarray}\label{eq:ep-condition}
    W_{\ell}(\omega_\star)=0\, ,\qquad \partial_\omega W_{\ell}(\omega_\star)=0\, ,\qquad\partial_\omega^2W_{\ell}(\omega_\star)\neq0 \, .
\end{eqnarray}
The double zero of $W_{\ell}$ gives a double pole of the Green function and hence a response linear in time. 
We also determine the EP numerically by solving the conditions in Eq.~\eqref{eq:ep-condition}, as detailed in Sec.~\ref{subsec:low-spectrum}.

We next introduce the particle source. At leading order in the mass ratio, the particle follows a timelike geodesic of the Schwarzschild background. By spherical symmetry we choose the equatorial plane, $z_{\text{p}}^\mu(t)=(t,r_{\text{p}}(t),\pi/2,\phi_{\text{p}}(t))$. 
For the ISCO separatrix plunge, the specific energy and angular momentum are $E=-u_t=2\sqrt{2}/3$ and $L=u_\phi=2\sqrt{3}M$, respectively \cite{Hadar:2009ip}. We choose $t=0$ and $\phi_{\text{p}}=0$ when the particle crosses $r_{\text{p}}=3M$, which fixes the time and azimuthal origins. The particle trajectory is then determined by
\begin{eqnarray}\label{eq:isco-plunge}
    \frac{\mathrm{d}r_{\text{p}}}{\mathrm{d}t} = -\frac{f(r_{\text{p}})}{2\sqrt{2}}\Big(\frac{6M}{r_{\text{p}}}-1\Big)^{3/2}\, , \qquad \phi_{\text{p}}(t) = 2\sqrt{3}\Bigg[1-\sqrt{\frac{r_{\text{p}}(t)}{6M-r_{\text{p}}(t)}}\Bigg]\, ,
\end{eqnarray}
with $2M<r_{\text{p}}<6M$. When extended backward in Schwarzschild time, this critical plunge geodesic asymptotically approaches the ISCO as $t\to-\infty$ ($r_{\text{p}}\to6M$).
The model adopted here is mainly intended to describe the plunge phase. To obtain the complete physical trajectory, one needs to account for the earlier orbital evolution, radiation reaction, and the inspiral-to-plunge transition \cite{Buonanno:2000ef,Ori:2000zn,Apte:2019txp}.

In the tortoise coordinate, the point particle source takes the form
\begin{eqnarray} \label{eq:source-x-main}
    S_{\ell m}(t,x)=a_{\ell m}(t)\delta[x-x_\text{p}(t)]+b_{\ell m}(t)\delta^{\prime}[x-x_\text{p}(t)]\, ,
\end{eqnarray}
where $x_\text{p}(t)=x[r_\text{p}(t)]$ and $\delta^{\prime}$ is the derivative of $\delta$ associated with $x$. The coefficients are given in Eq. \eqref{eq:ab_source_correction}. For the equatorial plunge, the axial source vanishes when $\ell+m$ is even \cite{Martel:2003jj}, as follows from the axial spherical harmonics in Eq. \eqref{eq:j0_general}. Thus, for $\ell=2$, the source vanishes for $m=0,\pm2$, and only $m=\pm1$ are excited. Since $\Psi_{2,-1}=-\Psi_{21}^{\star}$, we consider only the $(2,1)$ mode and do not include even-parity perturbations.
The waveform and spectral calculations below are restricted to $\ell=2$.

\begin{figure}[htbp]
 \centering
 \includegraphics[width=85mm]{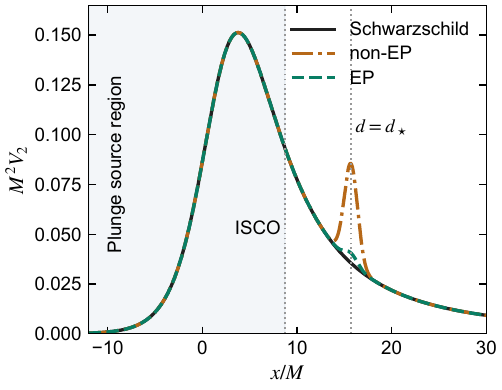}
 \caption{Axial $\ell=2$ potentials for Schwarzschild, the non-EP bump ($\epsilon=0.05$), and the EP potential ($\epsilon=\epsilon_\star=0.00508$), with $d=d_\star=15.6976M$ for both bumps. The shaded region is the plunge region $2M<r_\text{p}<6M$, which extends to $x\to-\infty$. Vertical lines mark the ISCO and the center of the bump.}
 \label{fig:physical-model}
\end{figure}

Figure \ref{fig:physical-model} compares the Schwarzschild potential with the EP potential at $(\epsilon_\star,d_\star)=(0.00508,15.6976M)$ and a non-EP potential at $(\epsilon,d)=(0.05,d_\star)$. The shaded region indicates the plunge region $2M<r_\text{p}<6M$, while the two vertical dotted lines mark the ISCO and the center of the Gaussian bump, respectively. The EP potential differs only slightly from the Schwarzschild potential, whereas the non-EP bump produces a more pronounced deviation.

\section{Frequency-domain Green function and QNM spectrum}
\label{sec:response}

\subsection{Homogeneous solutions and Green function}
\label{subsec:green}

We first construct the two homogeneous solutions $X_2^{\rm in}$ and $X_2^{\rm up}$ introduced in Sec. \ref{sec:theory}.  To remove their asymptotic exponential factors, we write
\begin{eqnarray}\label{eq:factored-solutions}
    X_2^{\rm in}(\omega,r)=\mathrm{e}^{-\mathrm{i}\omega x(r)}U^{\rm in}(\omega,r)\, ,\qquad X_2^{\rm up}(\omega,r)=\mathrm{e}^{\mathrm{i}\omega x(r)}U^{\rm up}(\omega,r)\, ,
\end{eqnarray}
so that the reduced functions $U^{\rm in/up}$ satisfy
\begin{eqnarray}\label{eq:factored-ode}
    f^2\frac{\mathrm{d}^2U^{\rm in/up}}{\mathrm{d}r^2}+\big(ff_{,r}\mp2\mathrm{i}\omega f\big)\frac{\mathrm{d}U^{\rm in/up}}{\mathrm{d}r}-V_2U^{\rm in/up}=0\, ,
\end{eqnarray}
where the minus and plus signs correspond to $X_2^{\rm in}$ and $X_2^{\rm up}$, respectively.  The functions $U^{\rm in}$ and $U^{\rm up}$ are regular at the horizon and at infinity, respectively, where they are expanded as
\begin{eqnarray}\label{eq:boundary-series}
    U^{\rm in}(\omega,r)=\sum_{k=0}^{N_\text{H}}a_k [f(r)]^k\, ,\qquad U^{\rm up}(\omega,r)=\sum_{k=0}^{N_\infty}b_k\Big(\frac{M}{r}\Big)^k \, .
\end{eqnarray}
We set $a_0=b_0=1$, which gives the unit amplitudes in Eq. \eqref{eq:jost}, and truncate the expansions at $N_\text{H}=12$ and $N_\infty=24$. Substitution into Eq. \eqref{eq:factored-ode} determines the remaining coefficients recursively. 

The ingoing solution is initialized at $r_\text{H}=2.01M$ and integrated outward by Eq. (\ref{eq:factored-ode}). For real frequencies $\omega\in\mathbb{R}$, the up solution is initialized at
\begin{eqnarray}\label{eq:real-outer-boundary}
    r_\text{R}=20M+M\max\Big(80,\frac{30}{|M\omega|}\Big)\, .
\end{eqnarray}
It is then integrated inward along the real axis. The frequency-dependent initial position $r_\text{R}$ keeps the far-zone expansion accurate at small $|\omega|$.

For complex frequencies with $\operatorname{Im}\omega<0$, the factor $\mathrm{e}^{\mathrm{i}\omega x}$ grows as $x\to+\infty$, and a direct inward integration of the outgoing solution from a large real radius is unstable. We therefore rotate the exterior part of the integration contour into the complex plane,
\begin{eqnarray}\label{eq:complex-ray}
    r(\xi)=R+\xi \mathrm{e}^{\mathrm{i}\theta}\, ,\qquad 0\leq\xi\leq\xi_{\max} \, .
\end{eqnarray}
Here $\xi$ parametrizes the complex ray, and we set $\theta=42^\circ$. The far-zone expansion is imposed at its outer end, $\xi=\xi_{\max}$, and the solution is integrated inward to $\xi=0$, where the contour returns to the real axis at $r=R=20M$. This provides accurate values of $X_2^{\rm up}$ and its derivative at $r=20M$ for complex frequencies.

For a given $\omega$, the solutions $U^{\rm in}$ and $U^{\rm up}$
obtained above determine the Wronskian,
\begin{eqnarray}\label{eq:wronskian-factored}
    W_2(\omega)=f\Big(U^{\rm in}\frac{\mathrm{d}U^{\rm up}}{\mathrm{d}r}-\frac{\mathrm{d}U^{\rm in}}{\mathrm{d}r}U^{\rm up}\Big)+2\mathrm{i}\omega U^{\rm in}U^{\rm up}\, .
\end{eqnarray}
Although the right-hand side depends explicitly on $r$, $W_2(\omega)$ is independent of $r$. Its numerical variation with radius therefore provides a check of the integration accuracy. The retarded Green function, defined by
\begin{eqnarray}\label{eq:green-equation}
    \Big(\partial_x^2+\omega^2-V_2\Big)g_2(\omega;x,x')=\delta(x-x')\, ,
\end{eqnarray}
with ingoing behavior at the horizon and outgoing behavior at infinity, is
\begin{eqnarray}\label{eq:green-frequency}
    g_2(\omega;x,x')=\frac{X_2^{\rm in}(\omega,x_<)X_2^{\rm up}(\omega,x_>)}{W_2(\omega)}\, ,
\end{eqnarray}
with $x_<=\min(x,x')$ and $x_>=\max(x,x')$.

\begin{figure}[htbp]
 \centering
 \includegraphics[width=178mm]{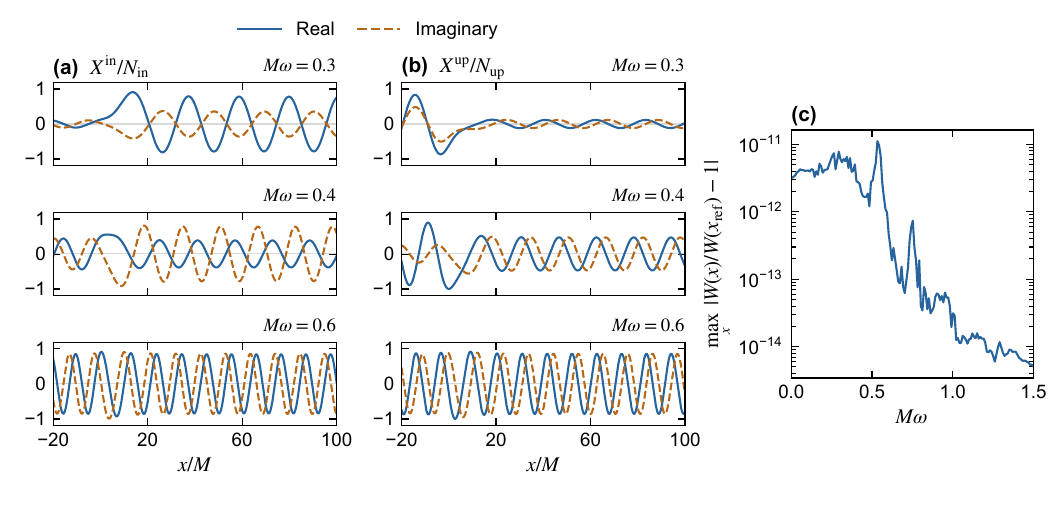}
\caption{Homogeneous solutions and a numerical consistency check for $(\epsilon_\star,d_\star)$. (a,b) Ingoing and outgoing solutions at three real frequencies, each divided by its maximum modulus $N_{\rm in/up}$ on the plotted range. Solid and dashed curves show real and imaginary parts. (c) Maximum relative variation of the Wronskian over $2.01\leq r/M\leq20$, with respect to its value at $x_{\rm ref}=x(4M)$.}
 \label{fig:homogeneous-validation}
\end{figure}

Figure \ref{fig:homogeneous-validation} shows the homogeneous solutions for the EP potential at $M\omega=0.3,0.4,0.6$, each normalized by its maximum modulus over the plotted range. As the frequency increases, their amplitude variations become less pronounced and the oscillations become more rapid. Panel (c) shows that the relative variation of the Wronskian remains small throughout the frequency range considered.

\subsection{Quasinormal modes and EP}
\label{subsec:low-spectrum}
Having obtained $W_2(\omega)$ for complex frequencies, we now determine the QNM spectrum by locating its zeros using complex Newton iteration. Since the iteration requires an initial guess sufficiently close to a root, we first locate the zeros through the phase winding of $W_2$. We consider the frequency region $2\times10^{-4}\leq\operatorname{Re}(M\omega)\leq2$ and $-0.45\leq\operatorname{Im}(M\omega)\leq2\times10^{-5}$, and divide it into rectangular cells. For each cell, the argument principle gives
\begin{eqnarray}\label{eq:winding-number}
    \nu_C=\frac{1}{2\pi \mathrm{i}}\oint_C\frac{\partial_\omega W_2(\omega)}{W_2(\omega)}\,\mathrm{d}\omega=\frac{1}{2\pi}\Delta_C\arg W_2=N_\text{Z}\, ,
\end{eqnarray}
where $C$ denotes the cell boundary and $\nu_C$ is the winding number of $W_2(\omega)$ around the origin, equal to the number $N_\text{Z}$ of zeros inside the cell, counted with multiplicity. The derivative $\partial_\omega W_2$ is obtained by differentiating the radial equations for $U^{\rm in/up}$ with respect to $\omega$ and solving simultaneously for $\partial_\omega U^{\rm in/up}$.  The cells with nonzero winding numbers identify their approximate locations. For each cell containing a simple zero ($N_\text{Z}=1$), we take its center as the initial guess and iterate
\begin{eqnarray}\label{eq:qnm-newton}
 \omega^{(k+1)}=\omega^{(k)}-\frac{W_2(\omega^{(k)})}{\partial_\omega W_2(\omega^{(k)})}\, .
\end{eqnarray}
The converged roots give the QNM spectra.

To determine the EP and the corresponding potential parameters, we instead impose $W_2(\omega_\star)=0$ and $\partial_\omega W_2(\omega_\star)=0$, and solve these two complex equations simultaneously for the four real variables $\operatorname{Re}\omega_\star$, $\operatorname{Im}\omega_\star$, $\epsilon_\star$, and $d_\star$. Starting from the values reported in Ref. \cite{Yang:2025dbn}, we solve the four real equations numerically using a \texttt{Python} root-finding routine and obtain
\begin{eqnarray}\label{eq:computed-ep}
    \epsilon_\star=0.00508294\, ,\qquad\frac{d_\star}{M}=15.6976\, ,\qquad M\omega_\star=0.365369-0.117441\mathrm{i}\, ,
\end{eqnarray}
which agrees with the values in Ref. \cite{Yang:2025dbn}. The winding number of the cell containing $\omega_\star$ is two, consistent with the coalescence of two QNMs. The nonzero $\partial_\omega^2W_2(\omega_\star)$ confirms that the zero is of second order.

\begin{figure}[htbp]
 \centering
 \includegraphics[width=85mm]
 {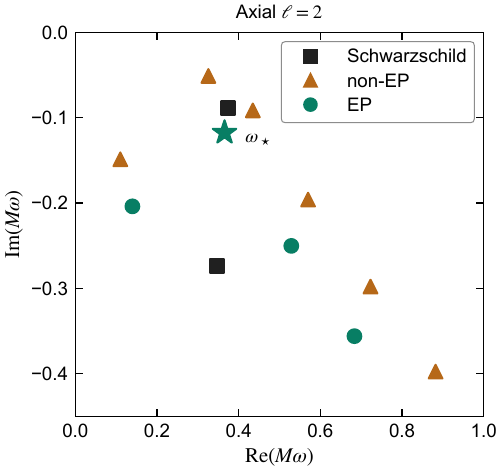}
 \caption{Axial $\ell=2$ QNMs of the three potentials in Fig. \ref{fig:physical-model}: Schwarzschild (squares), non-EP bump (triangles), and EP potential (green). The green star marks the degenerated spectrum $\omega_\star$. This mode is the least damped mode of the EP potential, while the remaining QNMs have larger damping rates.}
 \label{fig:low-spectrum}
\end{figure}

Figure \ref{fig:low-spectrum} shows the axial $\ell=2$ QNM spectra of the three potentials in Fig. \ref{fig:physical-model}, within $2\times10^{-4}\leq\operatorname{Re}(M\omega)\leq1$ and $-0.45\leq\operatorname{Im}(M\omega)\leq2\times10^{-5}$. Our search extends to $\operatorname{Re}(M\omega)=2$, but no additional QNMs are found beyond $\operatorname{Re}(M\omega)=1$. The winding-number calculation gives two, six, and five zeros for the Schwarzschild, non-EP, and EP potentials, respectively, counted with multiplicity, consistent with the results in Fig. \ref{fig:low-spectrum}. Comparing the modes with the smallest absolute imaginary parts in the three spectra, namely the fundamental ($n=0$) modes, we find $|\operatorname{Im}(M\omega_0^{\rm non-EP})|<|\operatorname{Im}(M\omega_0^{\rm Schw})|<|\operatorname{Im}(M\omega_\star)|$. The imaginary parts determine the exponential decay rates of these modes, which generally dominate the intermediate ringdown when sufficiently excited.

\section{Plunge waveform at the exceptional point}
\label{sec:results}
We now turn from the QNM spectrum to the waveform generated by the plunging particle. We first calculate the source-driven response and derive the EP contribution directly from the particle-source integral. We then compare the complete waveforms for the three potentials and identify the interval in which the EP contribution dominates. Finally, we examine its implications for QNM spectrum inference and Bayesian model selection.

\subsection{Particle-source response and EP waveform}
\label{subsec:particle-response}
We first calculate the frequency-domain response to the plunging particle using the Green function in Eq. \eqref{eq:green-frequency}. Fourier transforming Eq. \eqref{eq:master} gives $(\partial_x^2+\omega^2-V_2)\widetilde\Psi_{21} =\widetilde S_{21}$, where $\widetilde S_{21}(\omega,x') =\int \mathrm{d}t'\,\mathrm{e}^{\mathrm{i}\omega t'}S_{21}(t',x')$. The retarded solution is obtained by integrating the Green function against this source~\cite{Silva:2023cer},
\begin{eqnarray}
    \widetilde\Psi_{21}(\omega,x)=\int_{-\infty}^{\infty}\mathrm{d}x^{\prime}\,g_2(\omega;x,x')\widetilde S_{21}(\omega,x^{\prime})\, .
\end{eqnarray}
For an observer outside the entire plunge trajectory, we have $x_<=x^{\prime}$ and $x_>=x$ on the support of the source. Substituting Eqs. \eqref{eq:green-frequency} and
\eqref{eq:source-x-main} therefore gives
\begin{eqnarray}\label{eq:source-green-integral}
    \widetilde\Psi_{21}(\omega,x)=\frac{X_2^{\rm up}(\omega,x)}{W_2(\omega)}\int_{-\infty}^{\infty}\mathrm{d}t'\,\mathrm{e}^{\mathrm{i}\omega t'}\int_{-\infty}^{\infty}\mathrm{d}x'\,X_2^{\rm in}(\omega,x')\bigg\{a_{21}(t')\delta[x'-x_\text{p}(t')]+b_{21}(t')\delta^{\prime}[x'-x_\text{p}(t')]\bigg\}\, .
\end{eqnarray}
The delta-function term evaluates $X_2^{\rm in}$ along the particle trajectory. Integrating the derivative of the delta function by parts gives $-\partial_xX_2^{\rm in}$ there, with no boundary term. Since the up solution has unit amplitude, $X_2^{\rm up}(\omega,x)\sim \mathrm{e}^{\mathrm{i}\omega x}$ at infinity, the amplitude at infinity is~\cite{Silva:2023cer,Hadar:2009ip}
\begin{eqnarray}\label{eq:infinity-response}
    \widetilde\Psi_{21}^{\infty}(\omega)=\lim_{x\to\infty}\mathrm{e}^{-\mathrm{i}\omega x}\widetilde\Psi_{21}(\omega,x)=\frac{\mathcal J_{21}(\omega)}{W_2(\omega)}\, ,
\end{eqnarray}
where
\begin{eqnarray}\label{eq:excitation-integral}
    \mathcal J_{21}(\omega)=\int_{-\infty}^{\infty}\mathrm{d}t\,\mathrm{e}^{\mathrm{i}\omega t}\Big[a_{21}(t)X_2^{\rm in}(\omega,x_\text{p})-b_{21}(t)\partial_xX_2^{\rm in}(\omega,x_\text{p})\Big]\, .
\end{eqnarray}
Here $x_\text{p}=x_\text{p}(t)$. Combining the outgoing phase with the inverse Fourier kernel gives $\mathrm{e}^{-\mathrm{i}\omega t}\mathrm{e}^{\mathrm{i}\omega x}=\mathrm{e}^{-\mathrm{i}\omega u}$, with $u=t-x$.

The idealized ISCO-plunge geodesic extends to $t\to-\infty$ [see Eq.~\eqref{eq:isco-plunge}]. For the numerical calculation of the full waveform, we introduce a smooth window,
\begin{eqnarray}
    S_{21}^{\rm win}(t,x)=\chi(t)S_{21}(t,x)\, ,
\end{eqnarray}
where $\chi(t)=0$ for $t\leq-2400M$ and $\chi(t)=1$ for $t\geq-1800M$, with a smooth transition between them. The windowed source integral $\mathcal J_{21}^{\rm win}$ is obtained by multiplying the integrand in Eq.~\eqref{eq:excitation-integral} by $\chi(t)$. We then compute the waveform at infinity by inverse Fourier transformation,
\begin{eqnarray}\label{eq:infinity-waveform}
    \Psi_{21}^{\infty}(u)=\frac{1}{2\pi}\int_{-\infty}^{\infty}\mathrm{d}\omega\,\mathrm{e}^{-\mathrm{i}\omega u}\frac{\mathcal J_{21}^{\rm win}(\omega)}{W_2(\omega)}\, .
\end{eqnarray}

Equation \eqref{eq:infinity-waveform} gives the full particle-driven waveform for general values of the potential parameters $(\epsilon,d)$. We now fix $(\epsilon,d)=(\epsilon_\star,d_\star)$ to the EP values given in Eq.~\eqref{eq:computed-ep} and isolate the contribution from the double pole of the frequency-domain response. At the EP, $W_2(\omega_\star)=W_2^{\prime}(\omega_\star)=0$ and $W_2^{\prime\prime}(\omega_\star)\neq0$. We expand the frequency-domain response locally about $\omega=\omega_\star$. Since $W_2$ has a second-order zero there and the unwindowed $\mathcal{J}_{21}$ is analytic, this gives the Laurent expansion~\cite{Morikawa:2026fzn}
\begin{eqnarray}\label{eq:laurent-expansion}
    \frac{\mathcal J_{21}(\omega)}{W_2(\omega)}=\frac{C_{-2}}{(\omega-\omega_\star)^2}+\frac{C_{-1}}{\omega-\omega_\star}+\mathcal{O}(1)\, ,
\end{eqnarray}
where two constants $C_{-1}$ and $C_{-2}$ are
\begin{eqnarray}\label{eq:laurent-coefficients}
    C_{-2}=\frac{2\mathcal J_\star}{W_\star^{\prime\prime}}\, ,\qquad C_{-1}=\frac{2\mathcal J_\star'}{W_\star^{\prime\prime}}-\frac{2\mathcal J_\star W_\star^{\prime\prime\prime}}{3(W_\star^{\prime\prime})^2}\, .
\end{eqnarray}
Here $\mathcal J_\star$ and $W_\star$ denote the $\mathcal J_{21}(\omega)$ and the Wronskian $W_2(\omega)$ evaluated at $\omega=\omega_\star$, respectively, and primes denote derivatives with respect to $\omega$. To obtain the EP contribution, we evaluate the residue at $\omega=\omega_\star$ by expanding $\mathrm{e}^{-\mathrm{i}\omega u}$ in a Taylor series about $\omega_\star$ and combining the result with Eq. \eqref{eq:laurent-expansion}. This gives
\begin{eqnarray}\label{eq:ep-waveform}
    \Psi_{21}^{\rm EP}(u)=-\mathrm{i}\underset{\omega=\omega_\star}{\operatorname{Res}}\left[\mathrm{e}^{-\mathrm{i}\omega u}\frac{\mathcal J_{21}(\omega)}{W_2(\omega)}\right]=\big(A+Bu\big)\mathrm{e}^{-\mathrm{i}\omega_\star u}\, ,
\end{eqnarray}
with $A=-\mathrm{i}C_{-1}$ and $B=-C_{-2}$. The linear term is the characteristic time-domain signature of the EP~\cite{Yang:2025dbn}. Its coefficient is determined by the particle source and the Wronskian, rather than fitted to the waveform. In particular, the linear term is excited only when $\mathcal J_\star\neq0$. The mirror EP at $-(\omega_\star)^{\star}$ gives a separate contribution $\Psi_{21}^{\rm EP,mir}(u)$.

We denote by $H_{21}^{\rm odd}$ the odd-parity coefficient of the strain $r(h_+-\mathrm{i}h_\times)$ in its spin-weight $-2$ spherical-harmonic expansion. Reconstruction of the radiative metric at future null infinity relates this quantity to the RW function~\cite{Martel:2003jj,Martel:2005ir}, as also shown in Eq.~\eqref{eq:H21_Phi_app}. For the windowed source in Eq.~\eqref{eq:infinity-waveform}, we impose vanishing strain as $u\to-\infty$, yielding
\begin{eqnarray}\label{eq:strain}
    H_{21}^{\rm odd}(u)=-\mathrm{i}\sqrt{24}\int_{-\infty}^u\Psi_{21}^{\infty}(v)\,\mathrm{d}v \, ,
\end{eqnarray}
Substituting Eq.~\eqref{eq:ep-waveform} into Eq.~\eqref{eq:strain} gives
\begin{eqnarray}\label{eq:ep-strain}
    H_{21}^{\rm odd,EP}(u)=\frac{\sqrt{24}}{\omega_\star}\left(A+Bu-\frac{\mathrm{i}B}{\omega_\star}\right)\mathrm{e}^{-\mathrm{i}\omega_\star u}\, .
\end{eqnarray}
We decompose the complete strain multipole as
\begin{eqnarray}\label{eq:ep-decomposition}
    H_{21}^{\rm odd}(u)=H_{21}^{\rm odd,EP}(u)+H_{21}^{\rm odd,EP,mir}(u)+H_{21}^{\rm odd,rem}(u)\, ,
\end{eqnarray}
where the mirror term follows from Eq. \eqref{eq:ep-strain} with $\omega_\star$ replaced by $-(\omega_\star)^{\star}$ and $A$ and $B$ recomputed at that frequency. The term $H_{21}^{\rm odd,rem}$ contains the remaining contributions, including the prompt response, other QNMs, and the late-time tail.

\subsection{Waveform and EP-dominated ringdown}
\label{subsec:ep-waveform}
\begin{figure}[htbp]
 \centering
 \includegraphics[width=178mm]{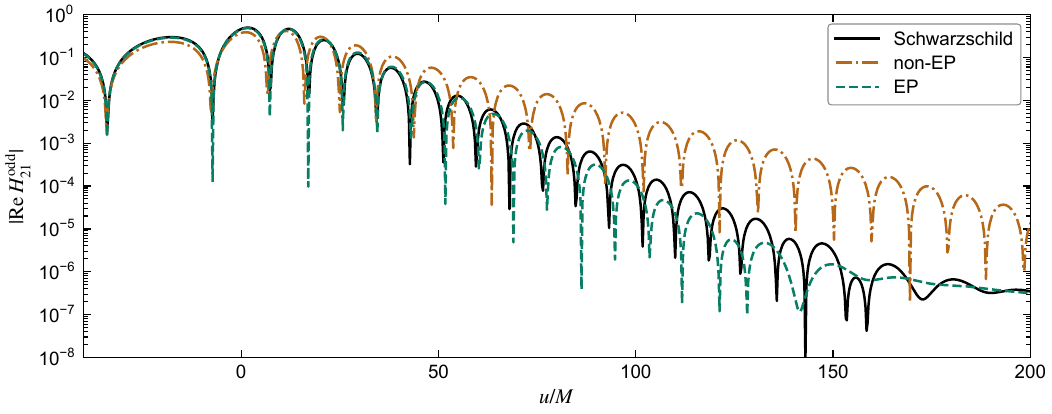}
 \caption{Plunge strains for Schwarzschild, the non-EP bump, and the EP potential, showing the absolute value of the real part. The horizontal axis shows retarded time $u$ in units of the black hole mass $M$.}
 \label{fig:plunge-waveform}
\end{figure}

We obtain the complete waveforms by numerically integrating Eq. \eqref{eq:infinity-waveform} along the real-frequency axis and converting the result to strain using Eq. \eqref{eq:strain}. Figure \ref{fig:plunge-waveform} compares the waveforms of the Schwarzschild, non-EP, and EP cases in Fig. \ref{fig:physical-model}. 
In the numerical calculations, we set $\mu=1$.

The three waveforms exhibit similar oscillations during the early plunge, reflecting their common particle trajectory and source, but differ more significantly during ringdown. The non-EP waveform has a longer oscillation period and slower decay than the Schwarzschild waveform, whereas the EP waveform has a similar oscillation period to Schwarzschild. These features are consistent with the QNM spectra in Fig. \ref{fig:low-spectrum}.

\begin{table}[htb]
 \centering
 \begin{tabular}{lcc}
 \hline\hline
 Pole & $A$ & $B$\\
 \hline
 $\omega_\star$
 & $0.014610+0.001562\mathrm{i}$
 & $0.001432+0.010265\mathrm{i}$\\
 $-(\omega_\star)^{\star}$
 & $0.002435-0.003560\mathrm{i}$
 & $-0.000652+0.000565\mathrm{i}$\\
 \hline\hline
 \end{tabular}
 \caption{Coefficients $A$ and $B$ in Eq. \eqref{eq:ep-waveform} for the EP at $\omega_\star$ and its mirror.}
 \label{tab:excitation}
\end{table}

\begin{figure}[!htbp]
 \centering
 \includegraphics[width=85mm]{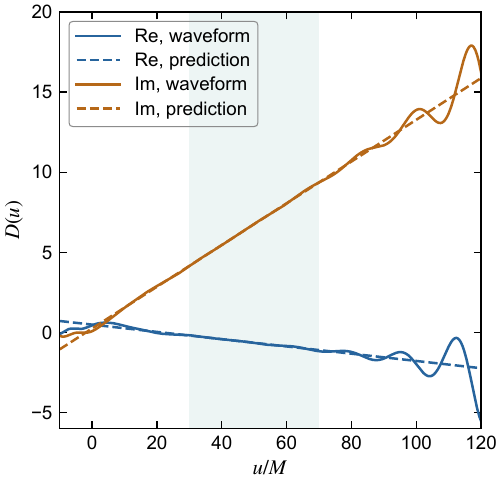}
 \caption{Real and imaginary parts of $D(u)$ (solid lines) and $A_\star+B_\star u$ (dashed lines), plotted against $u/M$. The shaded region marks $30\leq u/M\leq70$.}
 \label{fig:ep-prediction}
\end{figure}

We now examine the EP waveform during ringdown. Table~\ref{tab:excitation} lists the coefficients $A$ and $B$ in Eq.~\eqref{eq:ep-waveform} for the EP and its mirror. After subtracting the mirror contribution, we define
\begin{eqnarray}\label{eq:demodulated-strain}
    D(u)=\Big[H_{21}^{\rm odd}(u)-H_{21}^{\rm odd,EP,mir}(u)\Big]\mathrm{e}^{\mathrm{i}\omega_\star u}\, ,
\end{eqnarray}
Combining Eqs. \eqref{eq:ep-strain} and \eqref{eq:ep-decomposition} gives 
\begin{eqnarray}
    D(u)=A_\star+B_\star u+H_{21}^{\rm odd,rem}(u)\mathrm{e}^{\mathrm{i}\omega_\star u},
\end{eqnarray}where
\begin{eqnarray}
    A_\star=\frac{\sqrt{24}}{\omega_\star}\left(A-\frac{\mathrm{i}B}{\omega_\star}\right),\qquad
B_\star=\frac{\sqrt{24}}{\omega_\star}B\, .
\end{eqnarray}
Here $A$ and $B$ are the coefficients in the first row of Table~\ref{tab:excitation}. When the remaining response is small, $D(u)\simeq A_\star+B_\star u$.

The solid lines in Fig. \ref{fig:ep-prediction} show the real and imaginary parts of $D(u)$, obtained from the full numerical strain using Eq. \eqref{eq:demodulated-strain}. The dashed lines show the real and imaginary parts of $A_\star+B_\star u$, calculated from the coefficients in Table \ref{tab:excitation} using Eq. \eqref{eq:ep-strain}. They agree well over $30\leq u/M\leq70$, showing that the EP and mirror contributions dominate this stage of ringdown. We use this interval for the ringdown inference below.

\subsection{Ringdown inference}
\label{subsec:ringdown-inference}
We now investigate whether conventional ringdown templates can recover the underlying EP frequency from the complete particle-driven waveform~\cite{Yang:2025dbn,Macedo:2025ep,Cheng:2026gxu}. Here we use waveform fitting and Bayesian analysis to compare ringdown templates and assess frequency recovery. In conventional QNM analyses, the fitting interval is often chosen after the waveform peak to reduce the influence of early transients and rapidly damped overtones \cite{Isi:2021iql}. At sufficiently late times, however, the relative contribution of power-law tails increases \cite{DeAmicis:2024not}. We therefore focus on the intermediate ringdown interval $30\leq u/M\leq70$, beginning about $22M$ after the peak of the complex strain envelope at $u_{\rm peak}/M\simeq8.25$. As shown in Sec. \ref{subsec:ep-waveform}, the EP and mirror-EP contributions also dominate the waveform over this interval.

We compare an ordinary QNM template with a template containing the linear time dependence predicted at the EP. We also introduce a quadratic envelope to examine whether greater model flexibility necessarily improves frequency recovery as follows,
\begin{eqnarray}\label{eq:fitting-models}
    h_{\rm ord}(u) &=& c_+\mathrm{e}^{-\mathrm{i}\omega\tau}+c_-\mathrm{e}^{+\mathrm{i}\omega^{\star}\tau}\, ,\nonumber\\
    h_{\rm EP}(u) &=& (c_++b_+\tau)\mathrm{e}^{-\mathrm{i}\omega\tau}+(c_-+b_-\tau)\mathrm{e}^{+\mathrm{i}\omega^{\star}\tau}\, ,\nonumber\\
    h_{\rm quad}(u) &=& (c_++b_+\tau+d_+\tau^2)\mathrm{e}^{-\mathrm{i}\omega\tau}+(c_-+b_-\tau+d_-\tau^2)\mathrm{e}^{+\mathrm{i}\omega^{\star}\tau}\, ,
\end{eqnarray}
where $\tau=u-u_{\rm ref}$, with $u_{\rm ref}=30M$, and the amplitudes are independent complex parameters. The ordinary $h_{\rm ord}(u)$ and EP $h_{\rm EP}(u)$ templates are our primary models. Although multiple overtones can improve the description of early ringdown \cite{Isi:2021iql}, a single QNM often provides a useful approximation at intermediate times, after the more rapidly damped modes have decayed. We therefore investigate whether this conventional single-QNM description remains reliable in the presence of the EP resonance \cite{Yang:2025dbn}. The quadratic template $h_{\rm quad}(u)$ provides an additional test of whether improved fitting accuracy alone leads to a more accurate frequency estimate.

We fit all three templates to the complete complex strain. For each trial frequency $\omega$, the complex amplitudes are determined by linear least squares, while $\omega$ is optimized nonlinearly. We repeat the optimization from multiple initial frequencies and retain the solution with the smallest residual. The normalized waveform residual is
\begin{eqnarray}\label{eq:fit-residual}
    \mathcal{E}^2=\frac{\displaystyle \sum_{u_k\in\mathcal I} \Big|h_{\rm model}(u_k)-H^{\mathrm{odd}}_{21}(u_k)\Big|^2 }{ \displaystyle \sum_{u_k\in\mathcal I} \Big|H^{\mathrm{odd}}_{21}(u_k)\Big|^2 }\, , \qquad \mathcal I=\big[30M,70M\big]\, ,
\end{eqnarray}
where $h_{\rm model}$ denotes one of the three templates in Eq. \eqref{eq:fitting-models}.

The ordinary, EP, and quadratic templates contain six, ten, and fourteen real parameters, respectively. Since these models are nested, adding parameters cannot increase the minimum residual. However, a smaller residual $\mathcal{E}^2$ does not necessarily imply a more accurate QNM spectrum. We therefore compare both the fitting residual and the recovered frequency. As the relative errors in $\operatorname{Re}\omega$ are below $1\%$ for all three templates, we focus on the relative error in the imaginary part,
\begin{eqnarray}\label{eq:imag-error}
    \delta_{\mathrm{I}}=\Big( \frac{\operatorname{Im}\omega_{\rm fit}} {\operatorname{Im}\omega_\star}-1 \Big)\times100\%\, .
\end{eqnarray}
The results are summarized in Table \ref{tab:ringdown_templates}. The ordinary template gives a small waveform residual and underestimates the damping rate by $20.94\%$. The EP template improves the fit and reduces the damping-rate error to $0.08\%$. The quadratic template gives the smallest residual and overestimates the damping rate by $7.00\%$.

\begin{table}[htb]
 \centering
 \begin{tabular}{lcccc}
 \hline\hline
 Template & Real parameters & $M\omega_{\rm fit}$
 & $\mathcal{E}^2$ & $\delta_{\mathrm{I}}$ (\%)\\
 \hline
 Ordinary
 & 6
 & $0.36275-0.09285\mathrm{i}$
 & $1.845\times10^{-4}$
 & $-20.94$\\
 EP
 & 10
 & $0.36559-0.11735\mathrm{i}$
 & $2.024\times10^{-6}$
 & $-0.08$\\
 Quadratic
 & 14
 & $0.36197-0.12566\mathrm{i}$
 & $1.209\times10^{-6}$
 & $+7.00$\\
 \hline\hline
 \end{tabular}
 \caption{Fits to the complete EP-potential waveform on $30\leq u/M\leq70$. The independently calculated EP spectrum is $M\omega_\star=0.365369-0.117441\mathrm{i}$.}
 \label{tab:ringdown_templates}
\end{table}

The preceding fits reveal a systematic frequency bias, but do not show how it compares with the statistical uncertainty at finite SNR. We therefore perform a zero-noise injection~\cite{Isi:2021iql,Carullo:2019flw,Qiu:2023lwo}. We set the injected noise to zero at every sample, so the synthetic data are $d_k=H_{21}^{\rm odd}(u_k)$. The likelihood still assumes independent, zero-mean Gaussian noise with variance $\sigma_n^2>0$ in each real and imaginary component of each sample. This lets us assess template bias at finite SNR without fluctuations from an individual noise draw. The log-likelihood and SNR are
\begin{eqnarray}\label{eq:ringdown-likelihood}
    \ln\mathcal L=-\frac{1}{2\sigma_n^2} \sum_{u_k\in\mathcal I} \Big|d_k-h_{\rm model}(u_k)\Big|^2+\mathrm{const.}\, , \qquad \rho^2=\frac{ \displaystyle \sum_{u_k\in\mathcal I}|H_{21}^{\rm odd}(u_k)|^2 }{\sigma_n^2}\, ,
\end{eqnarray}
where $\mathcal I=[30M,70M]$. Here $\rho$ is the SNR assigned to this ringdown interval. We adopt uniform priors $\operatorname{Re}(M\omega)\in[0.15,0.65]$ and $\operatorname{Im}(M\omega)\in[-0.3,-0.015]$. For the amplitude parameters in Eq. \eqref{eq:fitting-models}, we adopt independent zero-mean Gaussian priors with width $a$ on their real and imaginary parts. We use $a=0.5$ as the reference width and also consider $a=0.25$ and $a=1$.

\begin{figure}[htbp]
 \centering
 \includegraphics[width=178mm]{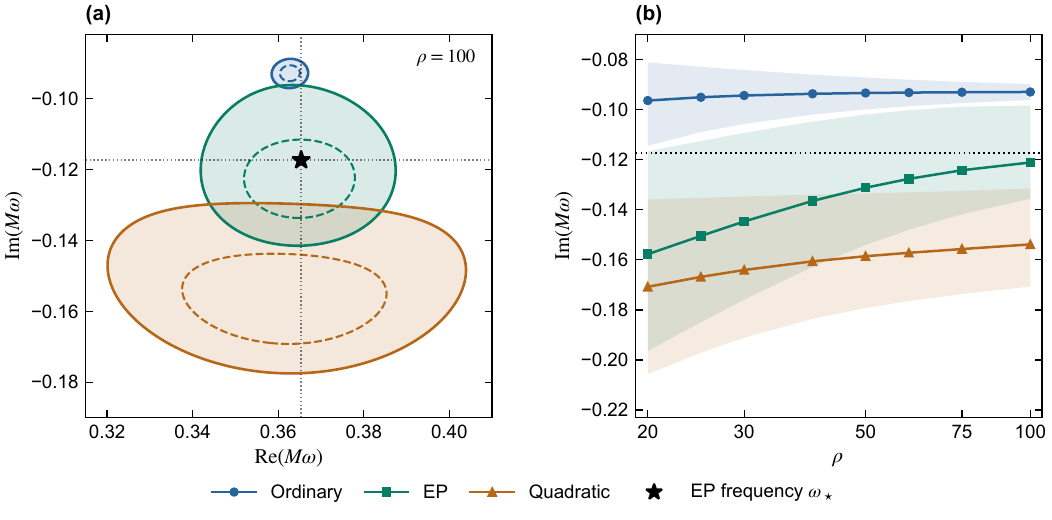}
    \caption{Frequency posteriors for the zero-noise injection on $30\leq u/M\leq70$ with $a=0.5$. (a) Joint posterior at $\rho=100$. (b) Posterior medians and equal-tailed $90\%$ credible intervals (shaded) for $\operatorname{Im}(M\omega)$ versus SNR. The black star in (a) marks the EP frequency $M\omega_{\star}$, and the dotted line in (b) marks its imaginary part.}
    \label{fig:plunge-fits}
\end{figure}

Figure \ref{fig:plunge-fits}(a) shows the joint posterior of the real and imaginary parts of $M\omega$ at $\rho=100$, after analytically marginalizing over the amplitudes. Dashed and solid contours enclose the $50\%$ and $90\%$ highest-posterior-density regions, respectively, and the black star marks the independently computed spectral value $M\omega_\star$. The ordinary template gives a narrow region that excludes the spectral value. The EP template gives a broader region that contains it, while the quadratic template gives the broadest region and still excludes it. Figure \ref{fig:plunge-fits}(b) shows the marginal posterior medians and $90\%$ credible intervals for $\operatorname{Im}(M\omega)$ as a function of SNR. At $\rho=20$, all three intervals are broad, and only the EP interval contains the spectral value. As $\rho$ increases, the intervals narrow and the medians tend to stabilize. Only the EP median approaches the spectral value closely. At $\rho=100$, the EP interval still contains the spectral value, while the ordinary and quadratic intervals continue to exclude it.

The ordinary template yields a narrow but biased frequency posterior. However, this does not determine which template is favored by the data. We therefore compare their Bayesian evidences,
\begin{eqnarray}
    Z_j=\int \mathcal L_j(\boldsymbol d\mid\boldsymbol\theta_j) \pi_j(\boldsymbol\theta_j) \,\mathrm{d}\boldsymbol\theta_j\, ,
\end{eqnarray}
Here $j\in\{\mathrm{ord},\mathrm{EP},\mathrm{quad}\}$ labels the template. For each template, $\boldsymbol\theta_j$ includes its frequency and amplitude parameters, $\mathcal L_j$ is the likelihood for the same injected data, and $\pi_j$ is the joint prior. We compare the EP template to the other two using $B_{\rm EP,ord}=Z_{\rm EP}/Z_{\rm ord}$ and $B_{\rm EP,quad}=Z_{\rm EP}/Z_{\rm quad}$. Unlike the minimum fitting residual, the Bayesian evidence averages the likelihood over the prior parameter space \cite{Trotta:2008qt}.

\begin{figure}[htbp]
 \centering
 \includegraphics[width=127.5mm]{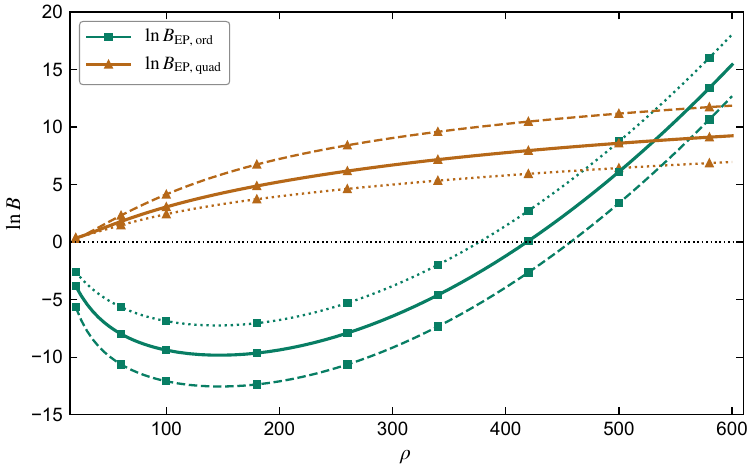}
    \caption{Log Bayes factors versus $\rho$. Squares mark $\ln(Z_{\rm EP}/Z_{\rm ord})$ and triangles mark $\ln(Z_{\rm EP}/Z_{\rm quad})$. Solid, dotted, and dashed curves use different widths $a=0.5$, $0.25$, and $1$, respectively. The horizontal dotted line marks equal evidence. Positive values favor the EP template.}
    \label{fig:plunge-bayes}
\end{figure}

Figure \ref{fig:plunge-bayes} shows the Bayes factors as functions of SNR for different $a$. At $\rho=100$ and $a=0.5$, we obtain $\ln B_{\rm EP,ord}=-9.39$ and $\ln B_{\rm EP,quad}=3.06$. The ordinary template is therefore favored over the EP template, despite yielding a substantially biased frequency estimate. At this SNR, the improvement in fit quality is insufficient to overcome the Occam penalty associated with the additional freedom of the EP template. At $\rho=500$, the corresponding log Bayes factors become $6.11$ and $8.59$, favoring the EP template over both alternatives. Although the Bayes factors depend quantitatively on the amplitude priors, the overall trend remains unchanged across the prior widths considered: the ordinary template is favored at lower SNR, whereas the EP template is favored at the higher SNRs. The qualitative conclusions from the frequency posterior at $\rho=100$ are also robust under these prior variations.

\section{Conclusions and discussion}
\label{sec:conclusions}
In this work, we studied the excitation of a second-order EP by a point particle plunging from the ISCO into a Schwarzschild black hole. We considered a Gaussian deformation of the axial RW potential and focused on the odd-parity mode $(\ell,m)=(2,1)$. Starting from a gravitational action with a curvature coupling, we derived the perturbation equation with the Gaussian bump modified potential and its point particle source, while the background geometry and leading-order particle trajectory remain those of Schwarzschild.

Using the frequency-domain Green function, we calculated the particle-driven waveform numerically and derived the EP contribution from the double pole of the response. This contribution has the form $(A+Bu)\mathrm{e}^{-\mathrm{i}\omega_\star u}$, with excitation coefficients determined directly from the particle source. The sum of the corresponding EP and mirror-EP strains agrees well with the numerical strain over $30\leq u/M\leq70$. This shows that the particle excites the characteristic EP response and that the EP and mirror contributions dominate the intermediate ringdown stage.

We then examine the implications for QNM spectra inference through waveform fitting and Bayesian analysis. The ordinary template gives a good waveform fit but underestimates the damping rate by about $21\%$, whereas the EP template reduces this error to below $0.1\%$. Adding a quadratic envelope further decreases the residual but yields a less accurate frequency than the EP template. At $\rho=100$, the ordinary template gives a narrow frequency posterior whose $90\%$ credible region excludes the true EP frequency, while the broader EP-template region contains it. Nevertheless, the Bayes factor favors the ordinary template at lower SNRs and the EP template only at higher SNRs. As said in the Introduction, our study provides an important warning for GW research: if we are unaware that the system is at an EP, a conventional template may fit the waveform well, yield a narrow posterior, and be favored by Bayesian evidence, but the inferred frequency is significantly biased.

However, further work is needed to connect this study more closely to gravitational-wave observations. Although we have realized the Gaussian bump modified potential from an action for $\ell=2$, extending this construction to other $\ell$ requires further investigation. Furthermore, we can investigate EPs in more general black hole spacetimes and their excitation by point particles following more complex trajectories. Using the resulting waveforms with realistic detector noise would then allow us to assess whether the inference bias found here remains significant in observational analyses.

\section*{Acknowledgments}
We thank for Li-Ming Cao for helpful discussion. This work is supported by the National Natural Science Foundation of China under Grant No. 12505067.

\appendix
\section{Lagrangian realization of the Gaussian bump modified potential}\label{app:reduced_action}

In this appendix, we derive a covariant model for the Gaussian potential and the point particle source used in the main text. The curvature coupling leaves the Schwarzschild background unchanged and modifies the axial perturbation equation~\cite{Endlich:2017tqa,Cardoso:2018ptl,Tattersall:2017erk}. We expand the action in the mass ratio $\mu/M$, reduce it to the odd-parity sector, and derive the sourced RW equation. We then determine the coupling denoted by $\mathcal{A}(\mathcal K)$ that gives the Gaussian bump for the axial quadrupole. We consider the gravitational action of Ref. \cite{Nakashi:2025fbr}, with a minimally coupled point particle,
\begin{eqnarray}\label{eq:full_action_app}
    I[g,z]=\frac{1}{16\pi}\int \mathrm{d}^4x\,\sqrt{-g} \Big[ R-\mathcal A(\mathcal K)\mathcal P^2 \Big]-\mu\int \mathrm{d}\lambda\, \sqrt{-g_{\alpha\beta}(z)\dot z^\alpha\dot z^\beta}\, ,
\end{eqnarray}
where $\mathcal K=R_{\alpha\beta\gamma\delta}R^{\alpha\beta\gamma\delta}$ is the Kretschmann scalar and $\mathcal P=\varepsilon^{\alpha\beta}{}_{\mu\nu}R^{\mu\nu\rho\sigma}R_{\alpha\beta\rho\sigma}$ is the Pontryagin pseudoscalar. We use signature $(-,+,+,+)$ and $\varepsilon_{tr\theta\phi}=\sqrt{-g}$, with no factor $1/2$ in the curvature dual. Both constant and curvature-dependent choices of $\mathcal A$, with $[\mathcal A]=\mathrm{length}^6$, have been considered in Refs. \cite{Endlich:2017tqa,Cardoso:2018ptl,Nakashi:2025fbr}. Here $\lambda$ is an arbitrary parameter along the particle trajectory and $\dot z^\mu\equiv\mathrm{d}z^\mu/\mathrm{d}\lambda$.

We expand the metric and the particle trajectory about a background $\bar g_{\mu\nu}$ in the mass ratio $\eta=\mu/M\ll1$,
\begin{eqnarray}\label{eq:metric_worldline_eta_expansion_app}
    g_{\mu\nu}=\bar g_{\mu\nu}+\eta h^{(1)}_{\mu\nu}+\eta^2h^{(2)}_{\mu\nu}+\mathcal{O}(\eta^3)\, ,\qquad z^\mu=z_0^\mu+\eta z_1^\mu+\mathcal{O}(\eta^2)\, .
\end{eqnarray}
The corresponding curvature invariants expand as
\begin{eqnarray}\label{eq:invariant_eta_expansion_app}
    R&=&\bar R+\eta R^{(1)}+\eta^2R^{(2)}+\mathcal{O}(\eta^3)\, ,\\
    \mathcal{K}&=&\overline{\mathcal K}+\eta\mathcal K^{(1)}+\eta^2\mathcal K^{(2)}+\mathcal{O}(\eta^3)\, ,\label{eq:K_eta_expansion_app}\\
    \mathcal{P}&=&\overline{\mathcal P}+\eta\mathcal P^{(1)}+\eta^2\mathcal P^{(2)}+\mathcal{O}(\eta^3)\, .\label{eq:P_eta_expansion_app}
\end{eqnarray}
The volume element is $\sqrt{-g}=\sqrt{-\bar g}\big(1+\eta s_1+\eta^2s_2\big)+\mathcal{O}(\eta^3)$, with
\begin{eqnarray}\label{eq:volume_coefficients_app}
    s_1=\frac12h^{(1)}\, ,\qquad
    s_2=\frac12h^{(2)}+\frac18\big(h^{(1)}\big)^2-\frac14h^{(1)}_{\mu\nu}h^{(1)\mu\nu}\, .
\end{eqnarray}
Here $h^{(n)}\equiv\bar g^{\mu\nu}h^{(n)}_{\mu\nu}$ and $h^{(n)\mu\nu}\equiv\bar g^{\mu\alpha}\bar g^{\nu\beta}h^{(n)}_{\alpha\beta}$. The inverse metric is $g^{\mu\nu}=\bar g^{\mu\nu}-\eta h^{(1)\mu\nu}+\mathcal{O}(\eta^2)$. The coupling function similarly gives
\begin{eqnarray}\label{eq:coupling_eta_expansion_app}
    \mathcal A(\mathcal K)=\overline{\mathcal A}+\eta\mathcal A^{(1)}+\eta^2\mathcal A^{(2)}+\mathcal{O}(\eta^3)\, ,\quad \text{with}\quad \mathcal A^{(1)}=\overline{\mathcal A}_{,\mathcal K}\mathcal K^{(1)}\, ,\qquad \mathcal A^{(2)}=\overline{\mathcal A}_{,\mathcal K}\mathcal K^{(2)}+\frac12\overline{\mathcal A}_{,\mathcal K\mathcal K}\big(\mathcal K^{(1)}\big)^2\, .
\end{eqnarray}
Here $\overline{\mathcal A}=\mathcal A(\overline{\mathcal K})$, and the derivatives of $\mathcal A$ are evaluated at $\overline{\mathcal K}$.
For the particle, $\bar\tau$ is proper time along $z_0^\mu$ in the background metric, with $\mathrm{d}\bar\tau=\sqrt{-\bar g_{\alpha\beta}(z_0)\dot z_0^\alpha\dot z_0^\beta}\,\mathrm{d}\lambda$. Choosing $\lambda=\bar\tau$, we define $u^\mu=\mathrm{d}z_0^\mu/\mathrm{d}\bar\tau$ and $a^\mu=u^\nu\bar\nabla_\nu u^\mu$, where $\bar\nabla$ is the background covariant derivative. Substituting Eq. \eqref{eq:metric_worldline_eta_expansion_app} and $\mu=\eta M$ into the particle term of Eq. \eqref{eq:full_action_app}, expanding, and integrating by parts gives
\begin{eqnarray}\label{eq:pp_eta_expansion_app}
    I_{\mathrm{pp}}=-\eta M\int \mathrm{d}\bar\tau+\eta^2 \Big[ \frac{M}{2}\int \mathrm{d}\bar\tau\, h^{(1)}_{\mu\nu}u^\mu u^\nu-M\int \mathrm{d}\bar\tau\,a_\mu z_1^\mu \Big]+\mathcal{O}(\eta^3)\, .
\end{eqnarray}
Substituting Eqs. \eqref{eq:invariant_eta_expansion_app}--\eqref{eq:pp_eta_expansion_app} into Eq. \eqref{eq:full_action_app} gives $I=I^{(0)}+\eta I^{(1)}+\eta^2I^{(2)}+\mathcal{O}(\eta^3)$, with
\begin{eqnarray}\label{eq:action_orders_app}
    I^{(0)}&=&\frac{1}{16\pi}\int\mathrm{d}^4x\,\sqrt{-\bar g}\Big[\bar R-\overline{\mathcal A}\,\overline{\mathcal P}^{\,2}\Big]\, ,\\
    I^{(1)}&=&\frac{1}{16\pi}\int\mathrm{d}^4x\,\sqrt{-\bar g}\Big[R^{(1)}+s_1\bar R
    -\big(\mathcal A^{(1)}+s_1\overline{\mathcal A}\big)\overline{\mathcal P}^{\,2}
    -2\overline{\mathcal A}\,\overline{\mathcal P}\mathcal P^{(1)}\Big]-M\int\mathrm{d}\bar\tau\, ,\label{eq:action_first_order_app}\\
    I^{(2)}&=&\frac{1}{16\pi}\int\mathrm{d}^4x\,\sqrt{-\bar g}\Big[R^{(2)}+s_1R^{(1)}+s_2\bar R-\overline{\mathcal A}\big(\mathcal P^{(1)}\big)^2\Big]\nonumber\\
    &&-\frac{1}{16\pi}\int\mathrm{d}^4x\,\sqrt{-\bar g}\,\overline{\mathcal P}\Big[2\overline{\mathcal A}\mathcal P^{(2)}+2\big(\mathcal A^{(1)}+s_1\overline{\mathcal A}\big)\mathcal P^{(1)}+\big(\mathcal A^{(2)}+s_1\mathcal A^{(1)}+s_2\overline{\mathcal A}\big)\overline{\mathcal P}\Big]\nonumber\\
    &&+M\int\mathrm{d}\bar\tau\,\Big[\frac12h^{(1)}_{\mu\nu}u^\mu u^\nu-a_\mu z_1^\mu\Big]\, .\label{eq:action_second_order_app}
\end{eqnarray}

We now specialize to the Schwarzschild background in Eq. \eqref{eq:schwarzschild}, for which
\begin{eqnarray}\label{eq:background_invariants_app}
    \bar R_{\mu\nu}=0\, ,\qquad \bar R=0\, ,\qquad \overline{\mathcal K}=\frac{48M^2}{r^6}\, ,\qquad \overline{\mathcal P}=0\, .
\end{eqnarray}
Since $\overline{\mathcal P}=0$, the curvature coupling leaves the background field equations unchanged, and Schwarzschild spacetime remains an exact solution. Varying Eq. \eqref{eq:action_first_order_app} with respect to $z_0^\mu$ gives
\begin{eqnarray}\label{eq:background_geodesic_app}
    a^\mu=u^\nu\bar\nabla_\nu u^\mu=0\, ,
\end{eqnarray}
so the leading-order trajectory $z_0^\mu$ is a timelike geodesic in Schwarzschild spacetime. Substituting Eqs. \eqref{eq:background_invariants_app} and \eqref{eq:background_geodesic_app} into Eq. \eqref{eq:action_second_order_app} gives,
\begin{eqnarray}\label{eq:leading_field_action_app}
    I^{(2)}=\frac{1}{16\pi}\int\mathrm{d}^4x\,\sqrt{-\bar g}\Big[R^{(2)}+s_1R^{(1)}-\overline{\mathcal A}\big(\mathcal P^{(1)}\big)^2\Big]+\frac12\int\mathrm{d}^4x\,\sqrt{-\bar g}\,T^{(1)\mu\nu}h^{(1)}_{\mu\nu}\, ,
\end{eqnarray}
in which the first-order stress-tensor coefficient is
\begin{eqnarray}\label{eq:point_particle_stress_app}
    T^{(1)\mu\nu}(x)=M\int \mathrm{d}\bar\tau\, \frac{u^\mu u^\nu}{\sqrt{-\bar g}}\, \delta^{(4)}\Big[x-z_0(\bar\tau)\Big]\, ,
\end{eqnarray}
where $\delta^{(4)}$ is the coordinate Dirac delta distribution and $T^{\mu\nu}=\eta T^{(1)\mu\nu}+\mathcal{O}(\eta^2)$. Equation \eqref{eq:point_particle_stress_app} recovers exactly the surviving particle term in Eq. \eqref{eq:action_second_order_app},
\begin{eqnarray}\label{eq:particle_stress_coupling_app}
    \frac12\int\mathrm{d}^4x\,\sqrt{-\bar g}\,T^{(1)\mu\nu}h^{(1)}_{\mu\nu}=\frac{M}{2}\int\mathrm{d}\bar\tau\,\mathrm{d}^4x\,h^{(1)}_{\mu\nu}(x)u^\mu u^\nu\delta^{(4)}\Big[x-z_0(\bar\tau)\Big]=\frac{M}{2}\int\mathrm{d}\bar\tau\,h^{(1)}_{\mu\nu}(z_0)u^\mu u^\nu\, .
\end{eqnarray}

We now reduce $I^{(2)}$ to the odd-parity sector. On a spherical background there is no odd-parity scalar harmonic, so only $h^{(1)}_{tA}$, $h^{(1)}_{rA}$, and $h^{(1)}_{AB}$ occur \cite{Martel:2005ir}. For modes $\ell\geq2$, we write
\begin{eqnarray}\label{eq:odd_harmonic_decomp}
    h^{(1)}_{tA}=\sum_{\ell\geq2,m}h_0^{\ell m}(t,r)X_A^{\ell m}\, ,\qquad h^{(1)}_{rA}=\sum_{\ell\geq2,m}h_1^{\ell m}(t,r)X_A^{\ell m}\, .
\end{eqnarray}
Here $m=-\ell,\ldots,\ell$ and $A,B\in\{\theta,\phi\}$. We use $X_A^{\ell m}=\epsilon_A{}^B D_BY_{\ell m}$, with $D_A$ the covariant derivative on the unit two-sphere and $\epsilon_{\theta\phi}=\sin\theta$. The angular components are $h^{(1)}_{AB}=\sum_{\ell\geq2,m}h_2^{\ell m}(t,r)X_{AB}^{\ell m}$, where $X_{AB}^{\ell m}=D_{(A}X_{B)}^{\ell m}$. For the action below, we use real orthonormal spherical harmonics, with $\int\mathrm{d}\Omega\,Y^2=1$ and $\int\mathrm{d}\Omega\,X_AX^A=\ell(\ell+1)$. For complex harmonics, these products become $Y^\star Y$ and $X_A^\star X^A$~\cite{Martel:2005ir}.
We impose RW gauge, $h_2=0$, and define the gauge-invariant combination
\begin{eqnarray}\label{eq:Xi_axial_def}
    \Xi\equiv\partial_t h_1-\partial_r h_0+\frac{2h_0}{r}\, .
\end{eqnarray}
Substituting Eq. \eqref{eq:odd_harmonic_decomp} into Eq. \eqref{eq:leading_field_action_app}, integrating over the angle part gives \cite{Martel:2005ir,Kobayashi:2012kh,Nakashi:2025fbr}
\begin{eqnarray}\label{eq:gravity_odd_action_app}
    I^{(2)}=\frac{\ell(\ell+1)}{16\pi}\int\mathrm{d}t\,\mathrm{d}r\Big[\frac{\beta_\ell}{2}\Xi^2+\frac{\lambda_\ell}{2fr^2}h_0^2-\frac{\lambda_\ell f}{2r^2}h_1^2\Big]+I_{\rm pp}^{(2)}\, ,
\end{eqnarray}
where $I_{\rm pp}^{(2)}$ is the second term of Eq. \eqref{eq:leading_field_action_app} and $\lambda_\ell\equiv(\ell-1)(\ell+2)$. The function $\beta_\ell(r)$ is 
\begin{eqnarray}\label{eq:beta_covariant_app}
    \beta_\ell(r)=1-\frac{4608M^2\ell(\ell+1)}{r^8}\mathcal A\Big(\frac{48M^2}{r^6}\Big)\, .
\end{eqnarray}
The calculations of $I_{\rm pp}^{(2)}$ are as follows. For the particle contribution, we use the equatorial Schwarzschild geodesic of Sec. \ref{sec:theory}, namely $u^t=E/f_\text{p}$, $u^\phi=L/r_\text{p}^2$, and $u^\theta=0$, where $f_\text{p}=f(r_\text{p})$ and $u^r=\mathrm{d}r_\text{p}/\mathrm{d}\bar\tau$. In RW gauge, only the symmetric $t\phi$ and $r\phi$ terms contribute to $h^{(1)}_{\mu\nu}u^\mu u^\nu$. Substituting Eq. \eqref{eq:odd_harmonic_decomp} and then using $\mathrm{d}\bar\tau=\mathrm{d}t/u^t$ gives
\begin{eqnarray}\label{eq:particle_mode_coupling_app}
    I_{\rm pp}^{(2)}&=&M\int\mathrm{d}\bar\tau\,u^\phi X_\phi\!\Big(\frac\pi2,\phi_\text{p}\Big)\Big[u^t h_0(t,r_\text{p})+u^r h_1(t,r_\text{p})\Big]\nonumber\\
    &=&M\int\mathrm{d}t\,\frac{L}{r_\text{p}^2}X_\phi\Big(\frac\pi2,\phi_\text{p}\Big)\Big[h_0(t,r_\text{p})+\frac{f_\text{p}u^r}{E}h_1(t,r_\text{p})\Big]\nonumber\\
    &=&M\int\mathrm{d}t\,\mathrm{d}r\,\frac{L}{r_\text{p}^2}X_\phi\Big(\frac\pi2,\phi_\text{p}\Big)\delta(r-r_\text{p})\Big[h_0(t,r)+\frac{f_\text{p}u^r}{E}h_1(t,r)\Big]\, .
\end{eqnarray}
The last line inserts the radial delta distribution to write the particle coupling as an integral over $t$ and $r$. Writing it as $\frac{\ell(\ell+1)}{16\pi}\int\mathrm{d}t\,\mathrm{d}r\,(j_0h_0+j_1h_1)$ defines the source coefficients
\begin{eqnarray}\label{eq:j0_general}
    j_0\equiv\frac{16\pi ML}{\ell(\ell+1)r_\text{p}^2}X_\phi\Big(\frac\pi2,\phi_\text{p}\Big)\delta(r-r_\text{p})\, ,\qquad j_1\equiv\frac{f_\text{p}u^r}{E}j_0\, .
\end{eqnarray}
Using these coefficients in Eq. \eqref{eq:gravity_odd_action_app} gives the complete quadratic action,
\begin{eqnarray}\label{eq:covariant_odd_action_app}
    I^{(2)}=\frac{\ell(\ell+1)}{16\pi}\int\mathrm{d}t\,\mathrm{d}r\Big[\frac{\beta_\ell}{2}\Xi^2+\frac{\lambda_\ell}{2fr^2}h_0^2-\frac{\lambda_\ell f}{2r^2}h_1^2+j_0h_0+j_1h_1\Big]\, .
\end{eqnarray}
The curvature coupling changes $\beta_\ell$ but leaves $j_0h_0+j_1h_1$ unchanged. For $\mathcal A=0$, $\beta_\ell=1$ and the Schwarzschild case is recovered. Varying Eq. \eqref{eq:covariant_odd_action_app} with respect to $h_1$ and $h_0$ gives
\begin{eqnarray}\label{eq:axial_constraints_app}
    -\partial_t(\beta_\ell\Xi)-\frac{\lambda_\ell f}{r^2}h_1+j_1=0\, ,\qquad \partial_r(\beta_\ell\Xi)+\frac{2\beta_\ell\Xi}{r}+\frac{\lambda_\ell}{fr^2}h_0+j_0=0\, .
\end{eqnarray}
Since the derivative terms involve $\beta_\ell\Xi$, we define the gauge-invariant master variable
\begin{eqnarray}\label{eq:Phi_def_app}
    \Phi\equiv\frac{r\beta_\ell\Xi}{\lambda_\ell}\, .
\end{eqnarray}
Solving Eq. \eqref{eq:axial_constraints_app} then gives
\begin{eqnarray}\label{eq:h0h1_Phi}
    h_0=-f\,\partial_r(r\Phi)-\frac{r^2f}{\lambda_\ell}j_0\, ,\qquad h_1=-\frac{r}{f}\partial_t\Phi+\frac{r^2}{\lambda_\ell f}j_1\, .
\end{eqnarray}
Substituting Eq. \eqref{eq:h0h1_Phi} into Eq. \eqref{eq:Xi_axial_def}, with $\Xi=\lambda_\ell\Phi/(r\beta_\ell)$, yields
\begin{eqnarray}\label{eq:Phi_master}
    \Big[-\partial_t^2+\partial_x^2-V_\ell^{(\beta)}(r)\Big]\Phi=-\frac{r}{\lambda_\ell}\Big[\partial_tj_1+f\partial_r(fj_0)\Big]\, ,
\end{eqnarray}
where $x$ is defined in Eq. \eqref{eq:tortoise} and
\begin{eqnarray}\label{eq:Vbeta_app}
    V_\ell^{(\beta)}=V_\ell^{\rm RW}+\frac{\lambda_\ell f}{r^2}\Big(\beta_\ell^{-1}-1\Big)\, .
\end{eqnarray}
The weak-coupling expansion agrees with Eq. (19) of Ref. \cite{Nakashi:2025fbr}. 

We define the main-text RW function as $\Psi_{\ell m}=-\eta f h_1^{\ell m}/r$. The physical RW function agrees with Eq. (A3) of Ref. \cite{Martel:2003jj}. Equation \eqref{eq:h0h1_Phi} then gives
\begin{eqnarray}\label{eq:PsiRW_Phi_relation}
    \Psi_{\ell m}=\eta\Big(\partial_t\Phi-\frac{rj_1}{\lambda_\ell}\Big)\, .
\end{eqnarray}
Applying the wave operator to Eq. \eqref{eq:PsiRW_Phi_relation} and using Eq. \eqref{eq:Phi_master} gives
\begin{eqnarray}\label{eq:modified_RW_source_app}
    \Big[-\partial_t^2+\partial_x^2-V_\ell^{(\beta)}(r)\Big]\Psi_{\ell m}=S_{\ell m}^{\rm Schw}+\eta\,\Delta V_\ell\frac{rj_1}{\lambda_\ell}\, ,\qquad \Delta V_\ell\equiv V_\ell^{(\beta)}-V_\ell^{\rm RW}\, ,
\end{eqnarray}
where the Schwarzschild source is~\cite{Martel:2003jj,Martel:2005ir}
\begin{eqnarray}\label{eq:RW_source_currents_app}
    S_{\ell m}^{\rm Schw}=-\frac{\eta}{\lambda_\ell}\Big[rf\,\partial_t\partial_r(fj_0)+\big(\partial_x^2-V_\ell^{\rm RW}\big)(rj_1)\Big]\, .
\end{eqnarray}
For the $(\ell,m)=(2,1)$ mode, the source coefficients in Eq. \eqref{eq:j0_general} uses $X_\phi^{21\star}(\pi/2,\phi_\text{p})=-\sqrt{15/(8\pi)}\,\mathrm{e}^{-\mathrm{i}\phi_\text{p}}$, giving
\begin{eqnarray}\label{eq:j21_currents}
    j_0^{21}=-\frac{2M\sqrt{30\pi}}{3}\frac{L}{r_\text{p}^2}\mathrm{e}^{-\mathrm{i}\phi_\text{p}}\delta(r-r_\text{p})\, ,\qquad j_1^{21}=\frac{f_\text{p}u^r}{E}\,j_0^{21}\, .
\end{eqnarray}
Substituting these into the full source on the right-hand side of Eq. \eqref{eq:modified_RW_source_app}, using the Schwarzschild geodesic equations and $\delta(r-r_\text{p})=f_\text{p}^{-1}\delta(x-x_\text{p})$, gives the coefficients in Eq. \eqref{eq:source-x-main},
\begin{eqnarray}\label{eq:ab_source_correction}
    a_{21}=N_{21}\mathrm{e}^{-\mathrm{i}\phi_\text{p}}\Big[\mathrm{i}\frac{L^2}{E}\frac{f_\text{p}^2}{r_\text{p}^4}-4\frac{L}{E}u^r\frac{f_\text{p}}{r_\text{p}^3}-\frac{Lu^r}{Er_\text{p}}\Delta V_2(x_\text{p})\Big]\, ,\qquad b_{21}=-\mathrm{i}N_{21}\frac{L^2}{E}\frac{f_\text{p}}{r_\text{p}^3}\mathrm{e}^{-\mathrm{i}\phi_\text{p}}\, .
\end{eqnarray}
Here $N_{21}=\mu\sqrt{30\pi}/6$ and $\mu=\eta M$. Setting $\Delta V_2=0$ recovers the Schwarzschild source.

We next choose the coupling $\mathcal{A}$ to reproduce the Gaussian modified potential in Eq. \eqref{eq:potential}. For $\ell=2$, $\lambda_2=4$, and the required potential correction is
\begin{eqnarray}\label{eq:target_gaussian_app}
    \Delta V_2=\frac{\epsilon}{M^2}\exp\Big[-\frac{(x(r)-d)^2}{M^2}\Big]\, .
\end{eqnarray}
Solving Eq. \eqref{eq:Vbeta_app} for $\beta_2$ gives
\begin{eqnarray}\label{eq:beta_gaussian_app}
    \beta_2(r)=\Bigg\{1+\frac{r^2}{4f(r)}\frac{\epsilon}{M^2}\exp\!\Big[-\frac{(x(r)-d)^2}{M^2}\Big]\Bigg\}^{-1}\, .
\end{eqnarray}
Then we can write
\begin{eqnarray}\label{eq:A_covariant_gaussian_app}
    \mathcal A(\mathcal K)=\frac{r^8}{27648M^2}\Big[1-\beta_2(r)\Big]\Bigg|_{r=(48M^2/\mathcal K)^{1/6}}\, ,\qquad 0<\mathcal K<\frac{3}{4M^4}\, .
\end{eqnarray}
For $\epsilon>0$, $0<\beta_2<1$ outside the horizon and $\beta_2\to1$ at the horizon and the infinity, so the master-variable transformation is regular.
This construction depends on $\ell$. With the coupling fixed by the $\ell=2$ potential,
\begin{eqnarray}\label{eq:beta_other_multipoles_app}
    \beta_\ell(r)=1-\frac{\ell(\ell+1)}{6}\Big[1-\beta_2(r)\Big]\, .
\end{eqnarray}
For sufficiently large $\ell$, $\beta_\ell$ crosses zero and becomes negative outside the horizon, making the potential singular and the coefficient of $\Xi^2$ negative. We therefore restrict this construction to $\ell=2$.

Finally, we derive the strain in Eq. \eqref{eq:strain}. At future null infinity, $j_0=j_1=0$ and $\beta_2\to1$. The outgoing solution of Eq. \eqref{eq:Phi_master} in this source-free region has $\Phi(t,r)=\Phi^\infty(u)+\mathcal O(r^{-1})$, with $u=t-x$. In RW gauge, $h_2=0$ and hence $h^{(1)}_{AB}=0$. The transformation to a radiation gauge with angular generator $\xi_A=\Lambda X_A$ is~\cite{Martel:2003jj,Martel:2005ir}
\begin{eqnarray}
    h_0^{\rm rad}=h_0^{\rm RW}-\partial_t\Lambda\, ,\qquad h_1^{\rm rad}=h_1^{\rm RW}-\partial_r\Lambda+\frac{2\Lambda}{r}\, ,\qquad h_2^{\rm rad}=h_2^{\rm RW}-2\Lambda\, .
\end{eqnarray}
Taking $\Lambda=r\Phi^\infty(u)+\mathcal O(1)$ gives
\begin{eqnarray}
    h_2^{\mathrm{rad}}=-2r\Phi^\infty(u)+\mathcal{O}(1)\, .
\end{eqnarray}
The gravitational-wave (GW) polarizations $h_+$ (plus) and $h_\times$ (cross) are defined by the transverse, trace-free radiative metric in the orthonormal angular basis $\mathrm{e}_{\hat\theta}=r^{-1}\partial_\theta$ and $\mathrm{e}_{\hat\phi}=(r\sin\theta)^{-1}\partial_\phi$, namely
\begin{eqnarray}
    \frac{\eta}{r^2}
    \begin{bmatrix}
        h^{(1),\rm rad}_{\theta\theta} & h^{(1),\rm rad}_{\theta\phi}/\sin\theta\\
        h^{(1),\rm rad}_{\theta\phi}/\sin\theta & h^{(1),\rm rad}_{\phi\phi}/\sin^2\theta
    \end{bmatrix}=\begin{bmatrix}h_+&h_\times\\h_\times&-h_+\end{bmatrix}\, .
\end{eqnarray}
The spin-weight $-2$ harmonic expansion of $h_+-\mathrm{i}h_\times$ defines $H_{2m}^{\rm odd}$ as follow,
\begin{eqnarray}\label{eq:strain_projection_app}
    h_+-\mathrm{i}h_\times=\frac{\eta}{r^2}\Big(h^{(1),\rm rad}_{\theta\theta}-\frac{\mathrm{i}}{\sin\theta}h^{(1),\rm rad}_{\theta\phi}\Big)=\frac1r\sum_{m=\pm1}H_{2m}^{\rm odd}(u)\,{}_{-2}Y_{2m}(\theta,\phi)+\mathcal O(r^{-2})\, .
\end{eqnarray}
Here, $H_{2m}^{\rm odd}$ has the $1/r$ distance dependence factored out. We now substitute $h^{(1),\rm rad}_{AB}=\sum_m h_2^{2m,\rm rad}X_{AB}^{2m}$. In our harmonic convention, $X_{\theta\theta}^{2m}-\mathrm{i}X_{\theta\phi}^{2m}/\sin\theta=(\mathrm{i}\sqrt{24}/2)\,{}_{-2}Y_{2m}$, which together with $h_2^{2m,\rm rad}=-2r\Phi_{2m}^\infty+\mathcal O(1)$ gives the first equality below. In the infinity, Eq. \eqref{eq:PsiRW_Phi_relation} reduces to $\Psi_{21}^\infty=\eta\,\partial_u\Phi_{21}^\infty$. Integrating gives Eq. \eqref{eq:strain}, i.e.,
\begin{eqnarray}\label{eq:H21_Phi_app}
    H_{21}^{\rm odd}(u)=-\mathrm{i}\eta\sqrt{24}\,\Phi_{21}^{\infty}(u)=-\mathrm{i}\sqrt{24}\int_{-\infty}^u\Psi_{21}^{\infty}(v)\,\mathrm{d}v\, .
\end{eqnarray}

\bibliography{reference}
\bibliographystyle{apsrev4-1}

\end{document}